\documentclass[aps, prl, amsmath, superscriptaddress,preprintnumbers,twocolumn]{revtex4-2}
\usepackage{graphicx}
\usepackage{dcolumn}
\usepackage{bm}
\usepackage{amsmath}
\usepackage{subfigure}
\usepackage{amsfonts}
\usepackage[svgnames]{xcolor}
\usepackage{longtable}
\usepackage{appendix}
\usepackage{multirow}
\usepackage{makecell}
\usepackage{booktabs}
\usepackage{tabularx}
\usepackage[colorlinks,
            linkcolor=blue,
			filecolor=blue,
			urlcolor= blue,
			citecolor=blue,
            ]{hyperref}
\usepackage{amsthm}
\usepackage{setspace}
\usepackage[T1]{fontenc}
\usepackage{extarrows}
\usepackage{lineno}

\usepackage[percent]{overpic}
\usepackage{dsfont}
\usepackage{floatrow}
\usepackage{physics}

\newcommand{\kett}[1]{\left.\ket{#1}\right\rangle}
\newcommand{\braa}[1]{\left\langle\bra{#1}\right.}
\newcommand{\brakett}[2]{\left\langle\braket{#1}{#2}\right\rangle}

\newcommand{\I}{\mathbb{I}}
\newcommand{\phii}[0]{\hat{\Phi}}
\newcommand{\mm}[0]{\hat{\mathcal{M}}}

\newcommand{\uu}[0]{\hat{\mathcal{U}}}

\renewcommand{\b}[0]{{b}}

\newcommand{\blue}[1]{\textcolor{blue}{#1}}

\def\mcc{\mathcal{C}}

\def\mcp{\mathcal{P}}

\begin{document}

\title{Resource-Efficient Noise Spectroscopy for Generic Quantum Dephasing Environments}
\author{Yuan-De Jin}\thanks{These authors contributed equally to this work}
\affiliation{State Key Laboratory of Semiconductor Physics and Chip Technologies, Institute of Semiconductors, Chinese Academy of Sciences, Beijing 100083, China}
\affiliation{Center of Materials Science and Opto-Electronic Technology, University of Chinese Academy of Sciences, Beijing 100049, China}

\author{Zheng-Fei Ye}\thanks{These authors contributed equally to this work}
\affiliation{Department of Applied Physics, University of Science and Technology Beijing, Beijing 100083, China}

\author{Wen-Long Ma}
\email{wenlongma@semi.ac.cn}
\affiliation{State Key Laboratory of Semiconductor Physics and Chip Technologies, Institute of Semiconductors, Chinese Academy of Sciences, Beijing 100083, China}
\affiliation{Center of Materials Science and Opto-Electronic Technology, University of Chinese Academy of Sciences, Beijing 100049, China}
\date{\today}
\begin{abstract}
We present a resource-efficient method based on repetitive weak measurements to directly measure the noise spectrum of a generic quantum environment that causes qubit phase decoherence. The weak measurement is induced by a Ramsey interferometry measurement (RIM) on the qubit and periodically applied during the free evolution of the environment. We prove that the measurement correlation of such repetitive RIMs approximately corresponds to
a direct sampling of the noise correlation function, thus enabling direct noise spectroscopy of the environment. Compared to dynamical-decoupling-based noise spectroscopy, this method can efficiently measure the full noise spectrum with the detected frequency range not limited by qubit coherence time. This method is also more resource-efficient than the correlation spectroscopy, as for the same detection accuracy with $N$ sampling times, it takes total detection time $O(N)$ while the latter one takes time $O(N^2)$. We numerically demonstrate this method for both bosonic and spin baths.


\end{abstract}

\maketitle

\textit{Introduction}---Knowledge of the environmental noise properties for a quantum system is essential for quantitatively understanding open-system dynamics \cite{Paladino2014,Suter2016} and designing robust-control or error-correcting strategies \cite{Viola1999,Faoro2004,Uhrig2007}. While longitudinal noises (causing energy relaxation) can be obtained by relaxometry \cite{Abragam1961,Schoelkopf2003,Kimmich2004,Tetienne2013,Pelliccione2014,Schmid-Lorch2015,Hall2016,Joas2017,Stark2017,Zhang2023,Heitzer2024}, the more dominant transerval noises (causing phase decoherence) are often charaterized by dynamical-decoupling- (DD) based noise spectroscopy \cite{Szańkowski2017,Cywiński2008,Yuge2011,Alvarez2011,Green2012,Paz-Silva2014,Soare2014,Kuffer2022}. The dephasing environment is called classical if it only imparts random
phases to superposition states of the qubit while suffering no backaction from the qubit, and called quantum if it is a genuine quantum system. The DD-based qubit noise spectroscopy infers the noise spectrum of a classical or quantum environment from qubit phase decoherence under a tailored sequence of $\pi$-pulses via the filter function formalism \cite{Cywiński2008,Yuge2011,Alvarez2011}. Over the past decade, this DD-based method has enabled the detection of environmental noise spectra in various physical systems \cite{Meriles2010,bylander2011,Yan2013,Kotler2013,Dial2013,Muhonen2014,Yoshihara2014,Myers2014,Romach2015,Chan2018,Romach2019,Farfurnik2020,Miln2021,Fu2021,Machado2023,Li2025,Shi2025}. However, its reliability is often limited by the probe qubit coherence time and pulse imperfections, leading to high sensitivity to pulse control errors, as well as the difficulty in resolving low-frequency noise. Moreover, this method is mostly conditioned on the Gaussian noise approximation \cite{Szańkowski2018}, so for complex quantum environments it may become unreliable to extract the noise spectra from the probe decoherence traces \cite{Ma2015,Yang2017,Hern2018}. Although several schemes have been proposed to partially overcome these limitations \cite{souza_robust_2012,Norris2016,Paz-Silva2017,Paz-Silva2019,Youssry2020, Chalermpusitarak2021,Maloney2022,Martina2023,Tripathi2024,Huang2025,Wang2025}, this method is still an indirect measurement strategy, so it can only detect a narrow range of noise spectrum for a specific DD sequence. 



An alternative method called correlation spectroscopy \cite{Laraoui2011,Laraoui2013,Kong2015,Staudacher2015,Zaiser2016,Rosskopf2017,Pfender2017} can directly reveal the full noise spectrum from temporal correlations of two probe readouts, avoiding complex probe control and probe coherence time limitation \cite{Young2012,Fink2013,Gefen2018,Sakuldee2020,Wudarski2023}, but existing analyses mostly treat classical environments and have not been extended to generic quantum environments \cite{Gefen2018,Sakuldee2020}. Moreover, it is unclear whether this method is resource-efficient (e.g., regarding the total detection time). On the other hand, weak measurement can monitor quantum dynamics while causing the least disturbance \cite{Aharonov1988,Korotkov2001a,Korotkov2001b,Wiseman2002,Jordan2005,Jordan2006}. In particular, sequential weak-measurement protocols have realized high-resolution sensing of single nuclear spins \cite{pfender2019,Cujia2019} and nuclear spin clusters \cite{Cujia2022}, characterization of high-order correlation functions \cite{Wang2019,Meinel2022,Wu2024} and nonlinear spectroscopy \cite{Vorobyov2023,Cheung2025} of quantum baths. However, because of the accumulated measurement backaction from probe to environment, it remains unknown whether the sequential weak-measurement strategy can reliably reconstruct the full noise spectrum of a quantum environment. These conceptual and practical challenges motivate the pursuit of a general framework for resource-efficient noise spectroscopy of a generic quantum environment. 


In this paper, we provide such a framework for efficient direct noise spectroscopy of generic quantum dephasing environments by repetitive weak measurements. Each weak measurement on the environment is realized by a Ramsey interferometry measurement (RIM) on the probe qubit and applied periodically during the free evolution of the environment. With the framework of sequential quantum channels, we prove that the measurement correlations between different RIMs approximately constitute a direct sampling of the noise correlation function, except that it differs from the exact one by a global factor and weak damping envelopes that cause slight broadening of the noise peaks. Therefore, our method can perform a direct measurement of the full noise spectrum, in contrast to the indirect measurement for DD-based noise spectroscopy. We also prove that the correlation spectroscopy initially developed for classical noises can
also perform noise spectroscopy of generic quantum environments. However, our method via repetitive weak measurements is more resource-efficient than the correlation spectroscopy, as for $N$ sampling times it takes total detection time $O(N)$ while the latter takes time $O(N^2)$. We numerically demonstrate the viability of this method for both bosonic and spin baths.

\begin{figure}[t]
    \centering
    \includegraphics[width=0.95\textwidth]{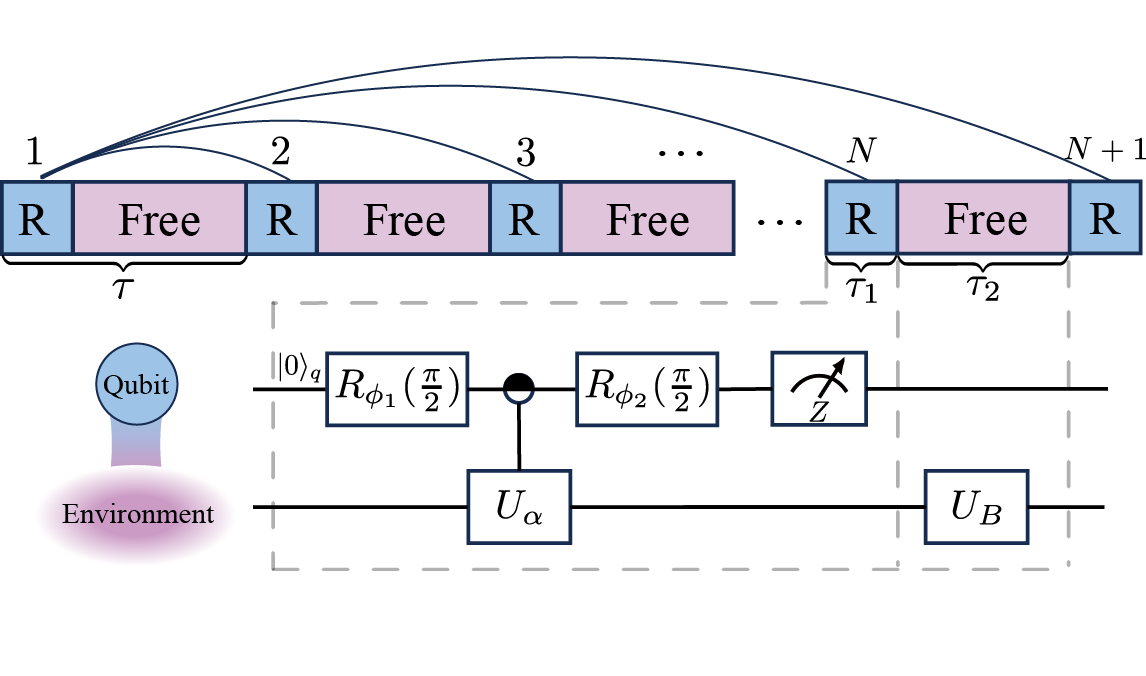} 
    \caption{Schematic of direct noise spectroscopy for a generic quantum dephasing environment via repetitive weak measurements. Each cycle contains a RIM on a probe qubit and a free evolution of the environment, denoted as "R" and "Free", respectively. When the RIM induces a weak measurement on the environment, the measurement correlations between the first and the other RIMs approximately constitute a direct sampling of the environmental noise correlation function.
    } 
    \label{fig:1} 
\end{figure}


\textit{Preliminaries}---We start by introducing the basic concepts of qubit noise spectroscopy. We consider a pure-dephasing Hamiltonian describing a qubit and its quantum environment
\begin{equation}\label{eq:Hamiltonian}
    H=\sigma^z_q\otimes A + \mathbb{I}_q\otimes B,
\end{equation}
where $\sigma_q^i$, $\mathbb{I}_q$ are the Pauli-$i$ ($i=x,y,z$) and identity operators of the qubit with $\sigma_q^z=|0\rangle_q\langle0|-|1\rangle_q\langle1|$,
$A$ is the noise operator for the qubit, $B=\sum_{j} b_j|j\rangle\langle j|$ is the free Hamiltonian of the environment.
Denoting the noise operator in the interaction picture as $ A_I(t)=e^{\,iBt}A\,e^{-iBt}$ and the environment state as $\rho$,
the symmetrized noise correlation function $\Tr\left\{\rho[  A_I(t) A_I(0)+ A_I(0) A_I(t)]\right\}/2$ is
\begin{equation}\label{Eq:ct}
        \mathcal C(t)=\sum_{i,j}|A_{ji}(A\rho)_{ij}|\cos (\omega_{ij}t+\varphi_{ij}),
\end{equation}
where $(\cdot)_{ij}=\langle i|(\cdot)|j\rangle$ denotes the matrix element in the basis of $B$, $\omega_{ij}=b_i-b_j$ and $\varphi_{ij}=\arg[A_{ji}(A\rho)_{ij}]$. Then the noise spectrum is defined as the spectral density function of $C(t)$, i.e., $S(\omega)=\int_{-\infty}^{\infty}dt\;e^{i\omega t}\,\mathcal C(t)=\pi\sum_{i,j}|A_{ji}(A\rho)_{ij}|\big[\delta(\omega-\omega_{ij})+\delta(\omega+\omega_{ij})\big]$.
Physically $S(\omega)$ quantifies the strength of environmental noise at frequency $\omega$ causing qubit phase fluctuations. 


Conventional DD-based noise spectroscopy obtains the spectrum through measuring the qubit decoherence under various DD sequences \cite{Cywiński2008,Yuge2011,Alvarez2011}. The simplest sequence is a RIM, during which the probe qubit with initial state $\ket{0}_q$ is first rotated to $\ket{\psi}_q=R_{\phi_1}(\frac{\pi}{2})\ket{0}_q=(\ket{0}_q-ie^{i\phi_1}\ket{1}_q)/\sqrt{2}$, with the rotation operator being $R_\phi(\theta)=e^{-i(\cos\phi\sigma_q^x+\sin\phi\sigma_q^y)\theta/2}$, then interacts with the environment under the Hamiltonian $H$ for time $t$, undergoes another rotation $R_{\phi_2}(\frac{\pi}{2})$, and is finally projectively measured in the basis $\{\ket{0}_q,\ket 1_q\}$.
A typical DD sequence employs a sequence of $N$ equally spaced $\pi$-pulses applied between two $\pi/2$ rotations, periodically flipping the qubit state to produce a filter function $f_t(t')$ that jumps between $-1$ and 1 at the flip pulses times. When the environment produces quantum Gaussian noise, the qubit coherence decays as
$W(t)=\exp(-\int_0^{\infty}\frac{d\omega}{2\pi}S(\omega) |{f}'_t(\omega)|^2 )$ with ${f}'_t(\omega)=\int_{-\infty}^{\infty}dt'\;e^{i\omega t'}f_t(t')$. So in the frequency domain, the DD pulse acts as a tunable filter function that selects specific spectral components of the noise. However, this method can only detect a narrow range of noise spectrum for a specific DD pulse sequence, and requires long probe coherent evolutions and accurate pulse control.   

\textit{Measuring the noise spectrum with repetitive RIMs}---We find that the noise spectrum can be efficiently extracted through correlation measurements of sequential RIMs during free evolution of the quantum environment (Fig.~\ref{fig:1}). This finding is based on exactly obtaining the measurement statistics of repetitive RIMs through the framework of sequential quantum channels.  

The RIM induces a quantum channel on the environment \cite{Jin2024,Qiu2024,Jin2025,Jin2025non}, which can be described in the Stinespring representation as $\Phi_R(\rho)=\Tr_q[ U(\ket{\psi}_q\bra{\psi}\otimes\rho) U^\dagger]$, with $U=e^{-iH \tau_1}=\sum_{\alpha=0}^1 |\alpha\rangle_q\langle \alpha|\otimes U_{\alpha}$ and $U_\alpha=e^{-i[(-1)^\alpha A+B]\tau_1}$. By partially tracing over the qubit, we obtain the Kraus representation of the channel $\Phi_R(\rho)=\sum_{a=0}^1M_a\rho M_a^\dagger$, where the Kraus operator is $M_a=[U_0-(-1)^a e^{i\Delta\phi}U_1]/2$ with $\Delta\phi=\phi_1-\phi_2$. Since the quantum channel is a superoperator on the $d$-dimensional Hilbert space, it can naturally be represented as a matrix on the $d^2$-dimensional vectorized operator space \cite{Watrous2018}. In such a space, an operator on the Hilbert space $A=\sum_{m,n=1}^dA_{mn}\ket{m}\bra{n}$ is vectorized as $\kett{A}=\sum_{m,n=1}^{d}A_{mn}\ket{m}\otimes\ket{n}$. Then a superoperator $X(\cdot)Y$ is equivalent to an operator $X\otimes Y^T$ on the vectorized operator space, and $\Phi_R$ can be represented as $\hat\Phi_R=\sum_{a=0}^1\mm_{a}$ with $\mm_{a}=M_a\otimes M_a^*$, where $(\cdot)^T$, $(\cdot)^*$ denote the matrix transposition and conjugation. 


After an RIM, the environment undergoes a free evolution with time $\tau_2$, which induces a unitary channel $\uu_{B}=U_{B}\otimes U_{B}^*$ with $U_{B}=e^{-iB\tau_2}$. 
The unitary evolution $U_B$ can be directly realized in platforms where the probe–environment interaction is tunable by external control (e.g., superconducting qubits \cite{Chen2014,Jones2013,Janzen2023} and trapped ions \cite{Wang2024a}) or preparing the probe in other idle states (e.g., nitrogen-vacancy (NV) center in a diamond \cite{Liu2017,pfender2019,Cujia2019}). 
For non-switchable probe–environment interaction, we can apply DD control to the probe during the free evolution to approximately realize $U_B$ (see Sec.~S5 of Supplementary Material (SM) 
\footnote{See the Supplementary Material 
for details about noise spectroscopy via repetitive weak measurements, correlation spectroscopy, the spin-boson and central spin models, effect of probe-environment coupling and performance analyses of different strategies for noise spectroscopy} for details). Then the concatenated channel is $\phii=\uu_{B}\phii_R$. If the concatenated channel $\phii$ is diagonalizable, we have $\hat{\Phi}=\sum_{k=1}^{d^2}\lambda_k\kett{R_k}\braa{L_k}$, where $\{\kett{R_k},\kett{L_k}\}$ forms a complete biorthonormal system satisfying $\brakett{L_j}{R_k}=\delta_{jk}$, with $\delta_{jk}$ being the the Kronecker delta. 


We relabel the measurement outcome of the $i$th RIM as $r_i\in\{+1,-1\}$ corresponding to $a_i\in\{0,1\}$. Then the function $\tilde C(m\tau)=\langle r_{m+1}r_1\rangle$, representing the two-point measurement correlation of the $(m+1)$th RIM and the first RIM, can be expressed as
\begin{equation}\label{Cm}
    \tilde C^{\rm weak}(m\tau)=\braa{\I}\hat\mcp\phii^{m-1}\hat\mcp\kett{\rho}=\sum_{k}d_{k}\lambda_{k}^{m-1},
\end{equation}
where $\hat\mcp=\uu_{B}(\mm_0-\mm_1)$, and the coefficient is $d_{k}=\langle\langle\I|\hat\mcp\kett{R_{k}}\braa{L_{k}}\hat\mcp\kett{\rho}$. For weak measurements, i.e., $\tau_1||A||\ll 1$ with $||\cdot||$ denoting the spectral norm, we can use perturbation theory to approximate the concatenated quantum channel (see Sec.~S1 of SM for details) \cite{Note1}
\begin{equation}\label{}
\phii\approx\hat{\mathcal{U}}_B'(\I+\tau_1^2\hat{\mathcal{L}}),
\end{equation}
where $\hat{\mathcal{U}}_B'=e^{-iB\tau}\otimes e^{iB^T\tau}=\sum_{ij}e^{-i\omega_{ij}\tau}\kett{ij}\braa{ij}$ with $\tau=\tau_1+\tau_2$, $\hat{\mathcal{L}}=-\hat{\mathcal{A}}^2$ and $\hat{\mathcal{A}}=A\otimes \I-\I\otimes A^T$. Then we can obtain the coefficients and eigenvalues in Eq.~\eqref{Cm} for $\Delta\phi=\pi/2$ as $d_{ij}\approx 2{\tau_1^2} A_{ji}(A\rho+\rho A)_{ij}e^{-i\omega_{ij}\tau}$, and $\lambda_{ij}=|\lambda_{ij}|e^{-i\omega_{ij}\tau}$ with $|\lambda_{ij}|=1+\tau_1^2\braa{ij}\hat{\mathcal{L}}\kett{ij}$.  
We thus derive the main result of this paper,
\begin{equation}
    \tilde C^{\rm weak}(m\tau)\approx4\tau_1^2\sum_{i,j}|A_{ji}(A\rho)_{ij}||\lambda_{ij}|^{m-1}\cos(m\omega_{ij}\tau+\varphi_{ij}),
\end{equation}
which differs from $\mathcal{C}(t)$ only by a global amplitude factor and weak damping envelopes that causes broadening of the noise peaks. So $\tilde C^{\rm weak}(m\tau)$ can be regarded as a direct sampling of the continuous-time function $\mcc(t)$ at time $t=m\tau$ with $m=1,\cdots,N$. 



We briefly summarize the applicability conditions of our method. First, the weak-measurement condition requires $\tau_1\ll 1/||A||$. Second, with the effective sampling period $\tau$ and $N$ sampling points, the detected frequency range is $[0,\pi/\tau]$ with spectral resolution $\pi/(N\tau)$, then we need $\omega_{\rm max}\leq 2||B||\leq \pi/\tau$, i.e., $\tau\leq\pi/(2||B||)$. Remarkably, our method has no requirement on the relative magnitude of $||A||$ and $||B||$ (see Sec.~S6 of SM) \cite{Note1}. Note that we have assumed that $A$ and $B$ are bounded operators on a finite-dimensional system. For  unbounded $A$ and $B$ on a infinite-dimensional system (such as a bosonic mode), the Hilbert space can often be truncated regarding the initial state to obtain effective bounded operators (see Sec.~S3 of SM) \cite{Note1}.

In simulations for practical experiments, the measurement correlation function is obtained from Monte Carlo samplings of quantum trajectories in repetitive quantum channels \cite{Jin2024}. After $N$ RIMs, the total channel can be decomposed into $2^N$ trajectories $ \phii^N=\sum_{r_1,r_2 \ldots, r_N=\pm 1} \hat{\mathcal{M}}_{r_N} \cdots \hat{\mathcal{M}}_{r_2}\hat{\mathcal{M}}_{r_1}$, with each trajectory labeled by the sequence of measurement outcomes $(r_1,r_2,...,r_N)$. 
Then the measurement correlation can be obtained by averaging over $N_s$ trajectories, i.e., $ \tilde C^{\rm weak}(m\tau)\approx\frac{1}{N_s}\sum_{i=1}^{N_s} r_1 r_m$.
According to the Hoeffding's inequality, $N_s$ should be at least $\frac{2}{\delta^2}\ln(\frac{2}{\epsilon}) $ to estimate each point within a fixed accuracy $\delta$ with probability $1-\epsilon$ (see Sec.~S6 of SM) \cite{Note1}. 

\begin{figure*}[t]
    \centering
    \includegraphics[width=18cm]{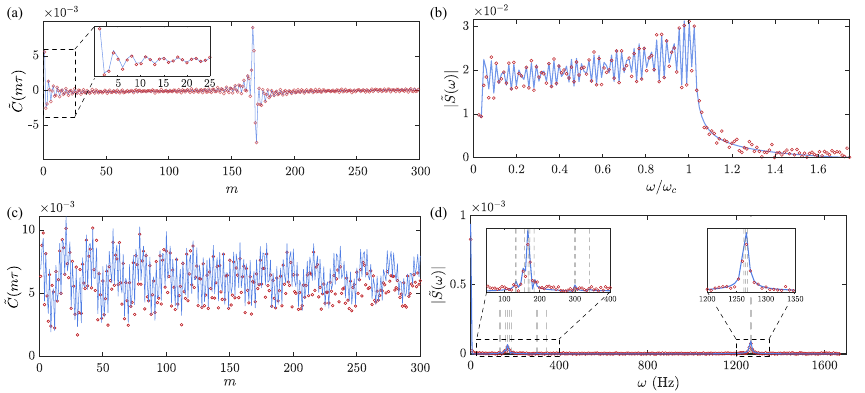}
    \caption{Direct qubit noise spectroscopy for (a,b) a bosonic bath and (c,d) a spin cluster bath. The blue solid lines represent the exact noise correlation functions (multiplied by a small global factor $4\tau_1^2$) in (a,c) and the corresponding noise spectra in (b,d), while the red circles represent the direct noise spectroscopy results obtained from Monte Carlo simulations of repetitive RIMs (with $4\times 10^6$ samples). For the Ohmic bosonic bath, we simulate only a finite set of discretized modes, and the residual tail in $|\tilde S(\omega)|$ arises from the finite-time discrete Fourier transform. The unitless parameters for the bosonic bath are $N_\omega=48$, $\alpha=0.4$, $\omega_{\rm max}=2$, $\tau_1=0.2$, $\tau=0.9$ , $\beta=1$, and the spin cluster contains five nuclear spins in the maximally mixed state with $\tau_1=1.8\,\,\mu$s, $\tau=300\,\,\mu$s, $B=0.1$ T and the leading theoretical frequencies $\omega_{ij}$ are indicated by dashed lines.}
    \label{fig:boson}
\end{figure*}

\begin{table*}[t]
\caption{Comparison of different strategies for qubit noise spectroscopy in fundamental limitations and total detection time. Here $N$ denotes the number of sampling times.} 
\centering 
\begin{ruledtabular}
\begin{tabular}{cccc}
Strategy & \makecell{Limited by Gaussian\\noise approximation} & \makecell{Limited by \\ probe coherence time}  & Detection time \\
\hline 
DD-based spectroscopy \cite{Cywiński2008,Yuge2011,Alvarez2011} & Yes & Yes & --- \\
Correlation spectroscopy \cite{Young2012,Fink2013,Gefen2018,Sakuldee2020} & No & No & $O(N^2)$ \\
Spectroscopy via repetitive weak measurements & No & No & $O(N)$ \\ 
\end{tabular}
\end{ruledtabular} \label{Table}
\end{table*}

\textit{Example I: Spin-boson model}---We first demonstrate our method for estimating the noise spectrum of a bosonic bath with a probe spin \cite{Reina2002,Uhrig2007}. The Hamiltonian of this model is $A=\sum_lg_l(\b^\dagger_l+\b_l)$ and $B=\sum_l \omega_l \b^\dagger_l \b_l$, where $g_l$ is the coupling strength between the probe and the $l$th bosonic mode, and $\b_l$ ($\b^\dagger_l$) is the annihilation (creation) operator of the $l$th bosonic mode. We assume that the bosonic bath is initially in the thermal state $\rho=e^{-\beta B}/\Tr(e^{-\beta B})$ with $\beta=1/(k_{\rm B}T)$ being the inverse temperature. 
Then the exact noise correlation function [Eq.~\eqref{Eq:ct}] and noise spectrum are $\mcc(t)=\sum_lg_l^2(2\bar n+1)\cos(\omega_l t)$ and $S(\omega)=\sum_l\pi g_l^2(2\bar n_l+1)[\delta(\omega-\omega_l)+\delta(\omega+\omega_l)]$, where $\bar n_l = 1/(e^{\beta\omega_l}-1)$. The bosonic bath can be characterized by a spectral density $J(\omega)=\sum_l g_l^2\delta(\omega-\omega_l)$. Here we choose the Ohmic bath with $J(\omega)=\alpha\omega\Theta(\omega_{\rm max}-\omega)$, where $\alpha$ is a dimensionless
parameter, $\Theta(x)$ is the step function and $\omega_{\rm max}$ is the cutoff frequency. We discretize the continuous spectrum into $N_\omega$ equally spaced frequencies with $\omega_l \in (0, \omega_{\rm max}]$, and
the simulation results in Fig.~\ref{fig:boson}\blue{(a-b)} show that our method can well recover the noise spectrum in the full frequency range (Sec.~S3 in SM \cite{Note1} for more simulations with different temperatures). 








\textit{Example II: Central spin model}---Then we apply our method for noise spectroscopy of a spin bath with a central spin \cite{Yuge2011}. We use the electron spin of an NV center (with $m_s$=$0,\pm 1$) to probe a dilute $^{13}$C nuclear spin bath in diamond. The Hamiltonian of the whole system in a strong magnetic field  \cite{Liu2017,deSousa2009,Ma2015}, with the NV electron in subspace $\{\ket{m_s=1}_q,\ket{m_s=-1}_q\}$, can be recast in the form of Eq.~\eqref{eq:Hamiltonian} with $A\approx-\sum_k{h}^z_k{{I}}_k^z$ and $B\approx-\omega_0 I_k^z+\sum_{j<k}D_{jk}[I_j^zI_k^z-(I_j^xI_k^x+I_j^yI_k^y)/2]$
, where $h_k^z$ is the vertical hyperfine interaction frequency of the $k$th nuclear spin, ${I}_k^{i}=\sigma_k^i/2$ $(i=x,y,z)$ is the $k$th nuclear spin operator, $\omega_0$ is the Larmor frequency of the nuclear spin and $D_{jk}$ denotes the strength of dipolar interaction between different nuclear spins. During the free evolution, the NV electron is reset to $\ket{m_s=0}_q$, then the hyperfine term is switched off and the nuclei evolve under the free Hamiltonian $B$.
We perform simulations for a cluster containing five ${}^{13}\text{C}$ nuclear spins, which are strongly coupled to each other. As shown in Fig.~\ref{fig:boson}\blue{(c-d)}, the noise spectroscopy results well reproduce the exact noise spectrum, revealing peaks centered at the theoretical coupling frequencies (see Sec.~S4 of SM for details) \cite{Note1}. We can also use $\{\ket{m_s=0}_q,\ket{m_s=-1}_q\}$ as the probe qubit, and perform direct noise spectroscopy by applying DD control during the free evolution to suppress the probe-bath coupling (Sec.~S4 in SM for simulations) \cite{Note1}.



\textit{Comparison with DD-based and correlation spectroscopy}---We compare our method based on repetitive weak measurements with DD-based and correlation spectroscopy regarding the fundamental limitations and consumed resources (Table \ref{Table}). DD-based noise spectroscopy is mostly effective when the Gaussian noise approximation and thus filter function formalism is valid. Although the filter formalism has been extended to non-Gaussian dephasing environments by engineering a multidimensional frequency comb, it is only applicable to classical environments and linearly coupled bosonic environments \cite{Norris2016}. Moreover, since a DD pulse with a specific frequency can only detect the noise spectrum near this frequency, scanning the DD pulse frequency with relatively large pulse numbers is necessary to accurately reconstruct the full noise spectrum, with the detected frequency range lower bounded by the inverse of qubit coherence time.




With the channel-based framework, we prove that the correlation spectroscopy method can also perform noise spectroscopy of generic quantum environments (see Sec.~S2 of SM) \cite{Note1}. The measurement correlation between two RIMs interleaved by an effective free evolution with duration $m\tau$ (including $\tau_1$ for the first RIM) is 
\begin{align}
\tilde \mcc^{\rm corr}(m\tau)&=\braa{\I}\hat \mcp(\uu_{B}')^{m-1}\hat\mcp\kett{\rho} \nonumber \\
 &\approx4\tau_1^2\sum_{i,j}|A_{ji}(A\rho)_{ij}|\cos(m\omega_{ij}\tau+\varphi_{ij}).
\end{align}
So the correlation spectroscopy, free of the measurement backaction during the free evolution, can more accurately reconstruct the noise spectrum than our repetitive weak-measurement method.
Note that the correlation spectroscopy also requires the weak-measurement conditions for the RIMs \cite{Note1}, resulting in signal amplitudes comparable to our method.



\begin{figure}[t]
    \centering
    \includegraphics[width=8.7cm]{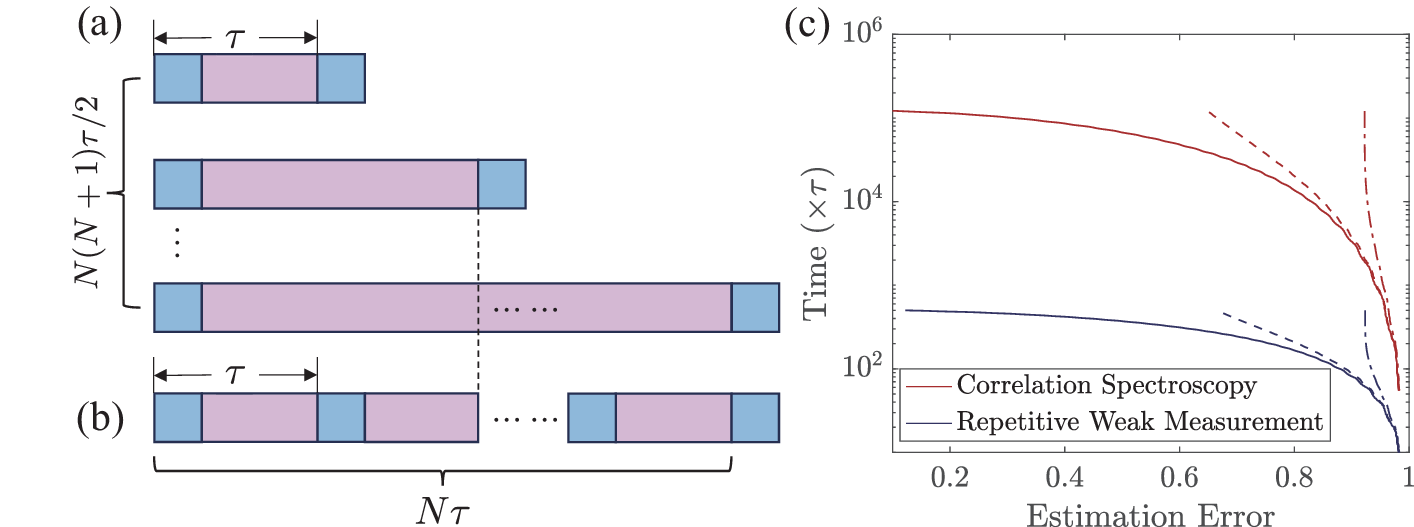} 
    \caption{Schematic illustration of the total detection time for (a) correlation spectroscopy and (b) noise spectroscopy via repetitive weak measurements. (c) Total detection time as a function of estimation error for different noise rates of the spin bath in Fig.~\ref{fig:boson}{\color{blue}(d)}: $\Gamma\tau=0$ (solid line), $\Gamma\tau=  10^{-3}$ (dashed line), and $\Gamma \tau =5\times10^{-3} $ (dash–dotted line), where we set $\Gamma_1=\Gamma_\phi=\Gamma$. The estimation error is defined as $||S-S^*||_2/||S||_2$, where $S$ is the ideal noise spectrum obtained from a long but finite-time evolution, $S^*$ is the noise spectrum obtained by different strategies, and $||\cdot||_2$ represents the Euclidean norm.} 
    \label{fig:2} 
\end{figure}

However, our method is more resource-efficient than the correlation spectroscopy in the total detection time. The correlation spectroscopy requires scanning the free evolution time [Fig.~\ref{fig:2}{\color{blue}(a)}], so when the set of evolution times are selected as $\{\tau,2\tau,\cdots,N\tau\}$, the total evolution time is $N(N+1)\tau/2$. Our method directly samples the correlation function at different times [Fig.~\ref{fig:2}{\color{blue}(b)}], requiring only the repetition of a fixed cycle, and reduce the total evolution time to $N\tau$. The improved scaling from $O(N^2)$ to $O(N)$ resembles the improved scaling from the standard to Heisenberg limit assisted by quantum coherence in quantum metrology \cite{Degen2017,Braun2018}. 

We numerically compare the performance of our method and correlation spectroscopy for a spin-bath environment [Fig.~\ref{fig:2}{\color{blue}(c)}]. For a noiseless spin bath, our method approximately achieve a square-root speedup for achievable estimation accuracy compared to correlation spectroscopy, while correlation spectroscopy can achieve higher accuracy with longer detection times. In practice, the spin bath suffers additional noises, including the bath spin relaxation errors $\sum_{k}\mathcal{D}[\sqrt{\Gamma_1}\sigma_k^-]$ and dephasing errors $\sum_{k}\mathcal{D}[\sqrt{\Gamma_\phi}\sigma_k^z]$, where $\mathcal{D}[\cdot]$ denotes the Lindblad dissipator, and $\sigma_k^-=(\sigma_k^x-i\sigma_k^y)/2$ with $\Gamma_1$ ($\Gamma_\phi)$ being the relaxation (dephasing) rate. Then the estimation accuracy of both methods tend to saturate as the total detection time increases. As the bath noise rates increase, the saturated accuracy of these two methods tends to coincide, so our method can still have advantage in reducing the detection time.  

\textit{Conclusions and outlooks}---We have proposed and analyzed a resource-efficient method for direct noise spectrospcpy of any quantum environment through repetitive weak measurements, implemented by repetitive RIMs on a probe qubit. Within the channel-based framework, we show that the two-point correlation function of the measurement statistics constitutes an approximate sampling of the quantum noise correlation, differing from the exact one only by a global amplitude factor and a weak damping envelope caused by weak measurements. Compared to DD-based noise spectroscopy, our method enables efficient measurement of the global noise spectrum while not limited by the probe lifetime and Gaussian noise approximation. It is also more resource-efficient than the correlation spectroscopy, as the scaling of detection time for $N$ sampling points is decreased from $O(N^2)$ to $O(N)$. 


Our method is applicable to a broad class of platforms with qubits coupled to quantum Gaussian or non-Gaussian environments with long memory times, such as solid-state defect spin qubits \cite{Laraoui2013,Hernandez2018,Bar2012}, superconducting qubits \cite{bylander2011,Ithier2005,Yan2013,Yoshihara2014}, trapped ions \cite{Kotler2013,Biercuk2009,Wei2022} and harmonic oscillators \cite{Milne2021,Keller2021}.
For quantum Gaussian noises, the reconstructed second-order noise correlation function can accurately predict the qubit phase decoherence dynamics, enabling more efficient design of robust control. For quantum non-Gaussian noises, the noise spectroscopy is still useful for characterization of quantum environments \cite{Rovny2024,Poteshman2025} and quantum sensing \cite{pfender2019,Cujia2019}.

By measuring the correlations between multiple RIM outcomes, our approach should enable efficient detection of high-order correlation functions \cite{Bednorz2013,Wang2019,DelRe2024}, which can used to characterize non-Gaussian quantum noises \cite{Sung2019,Wang2020,Kuffer2025,Xia2025,Dong2025}. It can also be applied to implement resource-efficient two-dimensional nuclear magnetic resonance (NMR) for resolving the connectivity in the noise spectrum \cite{Boss2016,Ma2016,Yang2020}. Additionally, it is interesting to extend the approach for noise spectroscopy with complex probes (e.g., a qudit \cite{Javaherian2025,Hesselmeier2024} or multiple qubits \cite{Szaifmmode2016,Lüpke2020,Sung2021,Wang2024,Rojas2025}), which can have enhanced spectral resolution and enable characterization of vectorial or correlated noises.


The research is supported by the National Natural Science Foundation of China (No. 12174379, No. 12574082, No. E31Q02BG), the Chinese Academy of Sciences (No. E0SEBB11, No. E27RBB11), Quantum Science and Technology-National Science and Technology Major Project (No. 2021ZD0302300) and Chinese Academy of Sciences Project for Young Scientists in Basic Research (YSBR-090).

\bibliography{ref,references}
\onecolumngrid 
\clearpage

\end{document}



\title{Supplementary Material for ``Resource-Efficient Noise Spectroscopy for Generic Quantum Dephasing Environments''}
\author{Yuan-De Jin}\thanks{These authors contributed equally to this work}
\affiliation{State Key Laboratory of Semiconductor Physics and Chip Technologies, Institute of Semiconductors, Chinese Academy of Sciences, Beijing 100083, China}
\affiliation{Center of Materials Science and Opto-Electronic Technology, University of Chinese Academy of Sciences, Beijing 100049, China}

\author{Zheng-Fei Ye}\thanks{These authors contributed equally to this work}
\affiliation{Department of Applied Physics, University of Science and Technology Beijing, Beijing 100083, China}

\author{Wen-Long Ma}
\email{wenlongma@semi.ac.cn}
\affiliation{State Key Laboratory of Semiconductor Physics and Chip Technologies, Institute of Semiconductors, Chinese Academy of Sciences, Beijing 100083, China}
\affiliation{Center of Materials Science and Opto-Electronic Technology, University of Chinese Academy of Sciences, Beijing 100049, China}
\date{\today }


\maketitle

\tableofcontents
 
\section{Details in noise spectroscopy via repetitive weak measurements}
In this section, we demonstrate the details of the method to estimate the noise spectrum through repetitive weak measurements, including the derivation of Eqs. (3) and (4) in the main text.

 In the end of each Ramsey interferometry measurement (RIM), the probe is projectively measured on the $Z$-basis $\{\ket{0}_q,\ket{1}_q\}$, while the outcomes are relabeled as $r\in\{+1,-1\}$ for convenience. 
Then the two-point correlation function is defined as 
\begin{equation}
    \tilde C^{\rm weak}(m)=\langle r_{m+1}r_1\rangle=\braa{\I}\hat\mcp\phii^{m-1}\hat \mcp\kett{\rho},
\end{equation}
where $r_m$ is the result of the $m$th RIM,  $\hat\mcp=\uu_{B}(\mm_0-\mm_1)$,  $\uu_{B}=U_{B}\otimes U_{B}^*$ is the unitary channel generated by $B$ with $U_{B}=e^{-iB\tau_2}$, and $\mm_a=M_a\otimes M_a^*$ is the Kraus superoperator of the channel induced by the RIM. As introduced in the main text, $M_a=[U_0-(-1)^a e^{i\Delta\phi}U_1]/2$ with $U_0=e^{-i(A+B)\tau_1}$ and $U_1=e^{-i(-A+B)\tau_1}$. We can spectrally decompose the quantum channel (i.e., $\phii=\sum_k\lambda_k\kett{R_k}\braa{L_k}$), then 
\begin{equation}\label{eq:cm}
\tilde C^{\rm weak}(m)=\sum_{k}\braa{\I}\hat\mcp\kett{R_k}\braa{L_k}\hat\mcp\kett{\rho}\lambda_k^{m-1},
\end{equation}
which is exactly Eq. (3) in the main text.

Then we analyze the eigenvalues of the concatenated channel. In the Hilbert-Schmidt (HS) space, the channel can be expressed as 
\begin{equation}
    \phii
    =\frac{1}{2}\uu_{B}\left(U_0\otimes U_0^*+U_1\otimes U_1^*\right),
\end{equation}
where $\uu_{B}=\sum_{ij}e^{-i\omega_{ij}\tau_2}\kett{ij}\braa{ij}$ with $\omega_{ij}=b_i-b_j$.
According to the Zassenhaus formula, up to the second order of $\tau_1$, we have
\begin{equation}\label{U0U1}
\begin{aligned}
    \begin{cases}
             U_0=e^{-iB\tau_1}\left[\mathbb{I}-i\tau_1 A-\frac{{\tau_1}^2}{2}(A^2+[A,B])\right]=e^{-iB\tau_1} \tilde U_0\\
             U_1=e^{-iB\tau_1}\left[\mathbb{I}+i\tau_1 A-\frac{{\tau_1}^2}{2}(A^2-[A,B])\right]=e^{-iB\tau_1} \tilde U_1.
    \end{cases}
\end{aligned}
\end{equation}
 we have 
\begin{equation}
    \tilde U_0\otimes \tilde U_0^*\approx\mathbb{I}\otimes \mathbb{I}-i\tau_1(A\otimes \mathbb{I}+\mathbb{I}\otimes A^T)+ \tau_1^2 \left[A\otimes A^T 
-\frac{1}{2}(A^2+[A,B])\otimes \mathbb{I} - \frac{1}{2} \mathbb{I}\otimes((A^T)^2 + [A^T,B^T])\right],
\end{equation}
and
\begin{equation}
   \tilde U_1\otimes \tilde U_1^*\approx\mathbb{I}\otimes \mathbb{I}+i\tau_1(A\otimes \mathbb{I}+\mathbb{I}\otimes A^T)+ \tau_1^2 \left[A\otimes A^T 
-\frac{1}{2}(A^2-[A,B])\otimes \mathbb{I} - \frac{1}{2} \mathbb{I}\otimes((A^T)^2 - [A^T,B^T])\right],
\end{equation}
resulting in
\begin{equation}\label{perturbChannel}
\begin{aligned}
        \phii&\approx\frac{1}{2}\hat{\mathcal{U}}_B'(\tilde U_0\otimes \tilde U_0^*+\tilde U_1\otimes \tilde U_1^*)
        =\hat{\mathcal{U}}_B'(\mathbb{I}+\tau_1^2\hat{\mcl}),
\end{aligned}
\end{equation}
where $\hat{\mathcal{U}}_B'=U_B'\otimes (U_B')^*=\sum_{ij}v_{ij}\kett{ij}\braa{ij}$ and $\hat \mcl=-(A\otimes \mathbb{I}-\mathbb{I}\otimes A^T)^2$ with $U_B'=e^{-iB\tau}$, $\tau=\tau_1+\tau_2$ and $v_{ij}=e^{-i\omega_{ij}\tau}$. We can use perturbation theory to obtain the eigenvalues of $\phii$,
\begin{equation}\label{lambdaper}
    \lambda_{ij}\approx v_{ij}+\tau_1^2\braa{ij}\uu_B\hat{\mcl}\kett{ij}=v_{ij}(1+\tau_1^2\braa{ij}\hat \mcl\kett{ij}).
\end{equation}
Since $\hat \mcl$ is a negative definite matrix, $\braa{ij}\hat \mcl\kett{ij}$ is a non-positive real number, we denote $|\lambda_{ij}|\approx 1+\braa{ij}\hat \mcl\kett{ij}\leq1$ and thus $\lambda_{ij}\approx v_{ij}|\lambda_{ij}|$.

We then consider the coefficients in Eq.~\eqref{eq:cm}. When $\tau_1||A||$ is small, the eigenstates of $\Phi$ are approximately unperturbed from $\uu_B$, which are $\{R_{ij}=L_{ij}=|i\rangle\langle j|\}_{i,j=1}^d$. Up to the first order, the Kraus operators of the concatenated channel are 
\begin{equation}\label{K0K1}
    \begin{aligned}
        \begin{cases}
            K_0\approx U_B' [(1-e^{i\Delta\phi})\I-i\tau_1 (1+e^{i\Delta\phi})A]/2,\\
            K_1\approx U_B' [(1+e^{i\Delta\phi})\I-i\tau_1 (1-e^{i\Delta\phi})A]/2.
        \end{cases}
    \end{aligned}
\end{equation}
Up to the first order of $\tau_1$,we have
\begin{equation}
\begin{aligned}
        \braa{\I}\hat\mcp\kett{ij}&=\Tr[(K_0^\dagger K_0-K_1^\dagger K_1)\ket{i}\bra{j}]\\
        &\approx \Tr\big[(-\cos(\Delta \phi) \I+2\tau_1\sin(\Delta\phi) A)\ket{i}\bra{j}\big]\\
        &=-\cos(\Delta\phi)\delta_{ij}+2\tau_1\sin(\Delta\phi) A_{ji},
\end{aligned}
\end{equation}
and 
\begin{equation}
\begin{aligned}
        \braa{ij}\hat\mcp\kett{\rho}&=\mel{i}{K_0\rho K_0^\dagger-K_1 \rho K_1^\dagger}{j}\\
        &\approx \mel{i}{U_B'\big[-\cos(\Delta\phi)\rho+\tau_1\sin(\Delta\phi)(A\rho+\rho A)\big]U_B'^\dagger}{j}\\
        &=e^{-i\omega_{ij}\tau}\big[-\cos(\Delta\phi)\rho_{ij}+\tau_1\sin(\Delta\phi)\sum_k (A_{ik}\rho_{kj}+\rho_{ik}A_{kj})\big],
\end{aligned}
\end{equation}
with $A_{ji}=\mel{j}{A}{i}$ and $\rho_{ij}=\mel{i}{\rho}{j}$. For $\Delta\phi=\pi/2$, we have 
\begin{equation}\label{Eq:ippr}
\begin{aligned}
        \braa{\I}\hat\mcp\kett{ij}\braa{ij}\hat\mcp\kett{\rho}
        &\approx 2{\tau_1^2}\sum_k A_{ji}(A_{ik}\rho_{kj}+\rho_{ik}A_{kj})e^{-i\omega_{ij}\tau},
\end{aligned}
\end{equation}
so the measurement correlation function is 
\begin{equation}
\begin{aligned}
    \tilde C^{\rm weak}(m)&\approx2{\tau_1^2}{}\sum_{i,j,k}A_{ji}(A_{ik}\rho_{kj}+\rho_{ik}A_{kj})|\lambda_{ij}|^{m-1}e^{-im\omega_{ij}\tau}\\
    &=2{\tau_1^2}\sum_{i,j,k}(A_{ji}A_{ik}\rho_{kj}|\lambda_{ij}|^{m-1}e^{-im\omega_{ij}\tau}+A_{ij}\rho_{jk}A_{ki}|\lambda_{ij}|^{m-1}e^{im\omega_{ij}\tau})\\
    &=4\tau_1^2\sum_{i,j}|A_{ji}(A\rho)_{ij}||\lambda_{ij}|^{m-1}\cos(m\omega_{ij}\tau+\varphi_{ij}).
\end{aligned}
\end{equation}
Then we can perform a discrete Fourier transformation for the measurement corelation function (here we assume $N\to\infty$)
\begin{equation}
     \tilde S^{\rm weak}(\omega)=2\tau_1^2\sum_{ij}|A_{ji}(A\rho)_{ij}|\left[\frac{e^{i\varphi_{ij}}}{1-|\lambda_{ij}|e^{-i(\omega-\omega_{ij})\tau}}+\frac{e^{-i\varphi_{ij}}}{1-|\lambda_{ij}|e^{-i(\omega+\omega_{ij})\tau}}\right],
\end{equation}
which suggests that the peaks are broadening as a result of the damping induced by the measurements.
%


\section{Details in correlation spectroscopy}
Here we briefly illustrate the method to obtain noise spectrum through the correlation spectroscopy. Each correlation measurement includes one initial polarization and readout (obtaining $r_I$) through RIM, a period of free evolution and a final RIM readout (obtaining $r_F$). Then the two-point correlation function of measurement results is 
\begin{equation}
    \tilde \mcc^{\rm corr}(m\tau)=\langle r_{F}r_I\rangle =\langle \I |\hat\mcp(\uu_B')^{m-1}\hat\mcp\kett{\rho},
\end{equation}
where $(\uu_B')^{m-1}$ is the unitary channel generated by free Hamiltonian $B$ with duration $t'=(m-1)\tau$. Then we can perform the spectral decomposition similarly, i.e., $\uu_B(t)=\sum_{ij} e^{-i\omega_{ij}t'}\kett{ij}\braa{ij}$, and insert it to the equation above
\begin{equation}
     \tilde\mcc^{\rm corr}(m\tau)=\sum_{ij}e^{-i\omega_{ij}t'}\langle \I |\hat\mcp\kett{ij}\braa{ij}\hat\mcp\kett{\rho}.
\end{equation}
In the weak-measurement limit, we can use Eq. \eqref{Eq:ippr}, and obtain
\begin{equation}
   \tilde \mcc^{\rm corr}(m\tau)\approx4\tau_1^2\sum_{i,j}|A_{ji}(A\rho)_{ij}|\cos(m\omega_{ij}\tau+\varphi_{ij})=4\tau_1^2\mcc(m\tau).
\end{equation}
We can see that the correlation function in this method is approximately the real noise correlation function without the damping induced by the sequential RIMs. However, we note that the correlation spectroscopy method still needs to work at the weak-measurement limit to recover the noise spectrum. Additionally, the total detection time of the correlation spectroscopy method is $t_{\rm tot}^{\rm corr}=\tau+2\tau+...+N\tau=N(N-1)\tau/2$, which significantly exceeds that of our method ($t_{\rm tot}^{\rm weak}=N\tau$).

\section{Details in the spin-boson model}

In this section, we derive the quantum channel for the spin-boson model, and show the feasibility of our method for different temperatures of the bosonic bath.

The Hamiltonian of the spin-boson model is
\begin{equation}
    H=\sum_l\left[\sigma_q^z\otimes g_l (\b^\dagger_l+\b_l)+\mathbb{I}_q\otimes \omega_l\b_l^\dagger \b_l \right] ,
\end{equation}
then we have
\begin{equation}
    A=\sum_{l}g_l (\b^\dagger_l+\b_l),\quad B=\sum_l\omega_l\b_l^\dagger \b_l.
\end{equation}
The bosonic bath can be characterized by a spectral density $J(\omega)=\sum_l g_l^2\delta(\omega-\omega_l)$ and it has been shown that the theoretical noise spectrum is
\begin{equation}
    S(\omega)=2\pi\sum_lg_l^2\left[\frac{1}{e^{\beta\omega_l}-1}+\frac{1}{2}\right][\delta(\omega+\omega_l)+\delta(\omega-\omega_l)],
\end{equation}
where $\beta=1/(k_{\rm B}T)$ and the initial state is taken as the thermal state $\rho=e^{-\beta B}/\Tr(e^{-\beta B})$. In the main text, we consider an Ohmic bath with $J(\omega)=\alpha\omega\Theta(\omega_{\rm max}-\omega)$ and $\omega_{\rm max}$ being the cutoff frequency. To discretize the continuous spectrum into $N_\omega$ discrete points, we have $ J(\omega_l)\Delta\omega=g_l^2$, i.e., $g_l=\sqrt{\alpha\omega_l\Delta\omega}\Theta(\omega_{\rm max}-\omega)$.

Since the multiple bosonic modes are independent, we can first consider a single bosonic mode, we denote $A_l=g_l (\b^\dagger_l+\b_l)$  and $B_l=\omega_l\b_l^\dagger \b_l=\sum_{n_l=0}^\infty n_l\omega_l\ket{n_l}\bra{n_l}$. Hereafter we omit the subscript $l$ when handling the single mode. 

We can analytically derive the exact quantum channel and show the validity of our results. We still consider the single mode bosonic mode first, we have
\begin{equation}
    A_I(t)=g (\b^\dagger e^{i\omega t}+\b e^{-i\omega t}),
\end{equation}
we note that the commutator
\begin{equation}
    [A_I(t_1),A_I(t_2)]=-2ig^2\sin[\omega(t_1-t_2)]
\end{equation}
is a $c$-number, thus
\begin{equation}
    U_0=e^{-i(A+B)\tau_1}=e^{-iB\tau_1}e^{\Omega_1+\Omega_2},
\end{equation}
in which
\begin{equation}
    \Omega_1=-i g\int_{0}^{\tau_1} (\b^\dagger e^{i\omega t}+\b e^{-i\omega t})\dd t=-\frac{g}{\omega}[\b^\dagger (e^{i\omega\tau_1}-1)-\b (e^{-i\omega \tau_1}-1)],
\end{equation}
and
\begin{equation}
    \Omega_2=ig^2\int_{0}^{\tau_1}\dd t_1\int_0^{t_1}\sin[\omega(t_1-t_2)]\dd t=i\frac{g^2}{\omega^2}(\omega\tau_1-\sin\omega\tau_1).
\end{equation}

Thus
\begin{equation}
    U_0=e^{-iB\tau_1}e^{-\frac{g}{\omega}[\b^\dagger \delta-\b \delta^*]}e^{\Omega_2},
\end{equation}
where $\delta=e^{i\omega\tau_1}-1\approx i\omega \tau_1$ when $\tau_1$ is small, and we note that $e^{\Omega_2}$ is still a $c-$number acting as a phase.
Similarly
\begin{equation}
    U_1=e^{-iB\tau_1}e^{\frac{g}{\omega}[\b^\dagger \delta-\b \delta^*]}e^{\Omega_2}.
\end{equation}

Let $D=\frac{g}{i\omega}[\b^\dagger \delta-\b \delta^*]$, then $U_\alpha=e^{-iB\tau_1}e^{-i(-1)^\alpha D}e^{\Omega_2}$. Thus
\begin{equation}
    \phii=\frac{1}{2}(U_{B}\otimes U_{B}^*)\sum_\alpha U_{\alpha}\otimes U_{\alpha}^*=\uu_B\cos(\hat \mcd),
\end{equation}
where $\hat\mcd=D\otimes \I+\I\otimes D^T$. When $\tau_1$ is small, 
\begin{equation}
    D\approx g\tau_1(b+b^\dagger)=\tau_1 A,
\end{equation}
then
\begin{equation}
    \cos(\hat\mcd)\approx1-\tau_1^2(A\otimes \I+\I\otimes A^T)^2,
\end{equation}
thus the form of channel coincides with  Eq. \eqref{perturbChannel}.



\begin{figure}
    \centering
    \includegraphics[width=\linewidth]{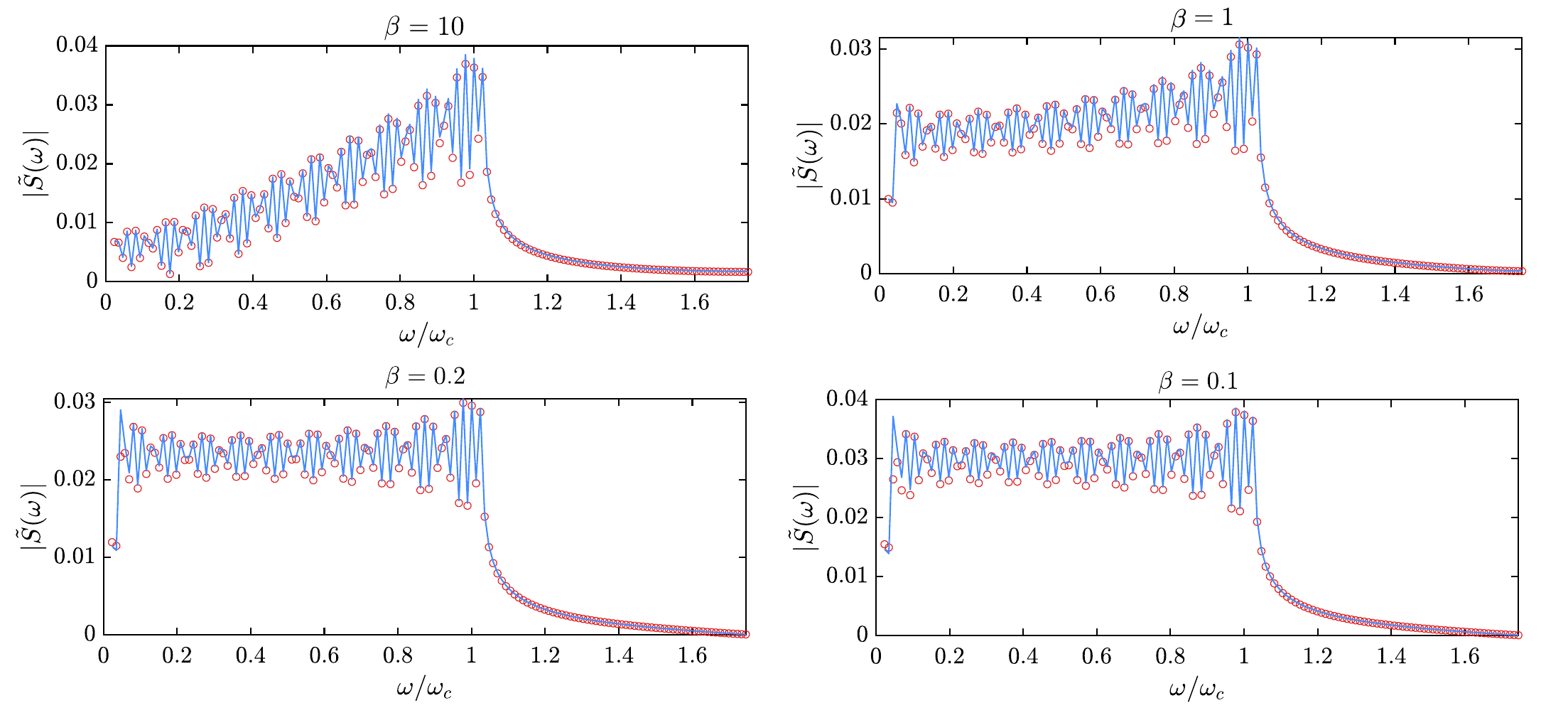}
    \caption{Direct qubit noise spectroscopy for a bosonic bath with different temperatures. The blue lines represent the ideal spectra and the red circles represent the direct noise spectroscopy results obtained from numerical matrix calculation. The unitless parameters are $\alpha=0.4$, $\omega_{\rm max}=2$, $\tau=0.9$, and $\tau_1=\{0.25,\,0.2,\,0.1,\,0.08\}$ for $\beta=\{10,\,1,\,0.2,\,0.1\}$.}
    \label{fig:spinbosonbeta}
\end{figure}

Below we show the spectroscopy for the spin-boson model in different temperatures in Fig.~\ref{fig:spinbosonbeta}. The figure shows that our method can detect the noise spectrum of the bosonic bath in a wide range of temperature.
Since the Hamiltonian $A$ is an unbounded operator, the Hilbert space should be truncated to define an effective norm $||A||_{\rm eff}$ regarding the initial state. Here we consider a thermal state as the initial state, and note that the thermal occupation number $\bar n = 1/(e^{\beta\omega}-1)$ increases with temperature. Then the truncated dimension of $A$ should be larger for a higher temperature, resulting in a larger effective norm $||A||_{\rm eff}$. So the coupling time $\tau_1$ is chosen to ensure a similar measurement strength $\tau_1||A||_{\rm eff}$.


\section{Details in the central spin model}\label{Sec:centralSpin}
Here we derive the noise spectrum of the $^{13}$C spin bath in the NV system. Under the secular approximation, the Hamiltonian of the $S=1$ NV electron spin and the $^{13}$C bath is
\begin{equation}
    H=S_z\otimes \sum_k\vb*{h}_k\cdot{\vb*{I}}_k-\omega_0\sum_k I_k^z+\sum_{j<k}D_{jk}'\big[{\vb*{I}}_j\cdot{\vb*{I}}_k-3({\vb*{I}}_j\cdot\vb*{r}_{jk})({\vb*{I}}_k\cdot\vb*{r}_{jk})/r_{jk}^2\big],
\end{equation}
where $S_z=P_{+1}+P_0-P_{-1}$ is the spin-1 operator with $P_{s}=\ket{m_s=s}_q\bra{m_s=s}$, $\vb*{h}_k=(h_k^x,h_k^y,h_k^z) $ is the hyperfine interaction vector of the $k$th nuclear spin, ${\vb*{I}}_k=(I_k^x,I_k^y,I_k^z)$ is the $k$th nuclear spin vector, $\vb*{r}_{jk}$ is the displacement from the $j$th target spin to the $k$th target spin, $\omega$ is the Larmor frequency of the nuclear spin and $D_{jk}'=\frac{\mu_0}{4\pi}\frac{\gamma_n^2\hbar^2}{r_{jk}^3}$ denotes the strength of dipolar interaction between different nuclear spins. We note that the $z-$axis is set to the direction of the magnetic field. 

\subsection{Noise spectroscopy of nearly independent nuclear spins}
We first analyze the case of nearly independent nuclear spins, where the bath can be decomposed into single-spin contributions, and the effective Larmor frequencies can be determined in the individual peaks in the noise spectrum. Then we can analytically derive the theoretical noise spectrum.
We can select the subspace $\{\ket{m_s=0}_q,\ket{m_s=-1}_q\}$, then the Hamiltonian can be recast to
\begin{equation}\label{Eq:Ham0-1}
    H=-(\sigma_q^z+\mathbb{I}_q)/2\otimes \sum_k\vb*{h}_k\cdot{\vb*{I}}_k-\omega_0\sum_k I_k^z+\sum_{j<k}D_{jk}\big[{\vb*{I}}_j\cdot{\vb*{I}}_k-3({\vb*{I}}_j\cdot\vb*{r}_{jk})({\vb*{I}}_k\cdot\vb*{r}_{jk})/r_{jk}^2\big],
\end{equation}
and thus
\begin{equation}
    A=-\frac{1}{2}\sum_k\vb*{h}_k\cdot{\vb*{I}}_k,\,\,B=-\frac{1}{2}\sum_k(\vb*{h}_k\cdot{\vb*{I}}_k+2\omega_0 I_k^z)+\sum_{j<k}D_{jk}\big[{\vb*{I}}_j\cdot{\vb*{I}}_k-3({\vb*{I}}_j\cdot\vb*{r}_{jk})({\vb*{I}}_k\cdot\vb*{r}_{jk})/r_{jk}^2\big],
\end{equation}
Since the nuclear spins are nearly independent, i.e., $D_{jk}\ll h_k,\omega$, the Hamiltonian can be approximated as
\begin{equation}
     A=-\frac{1}{2}\sum_k\vb*{h}_k\cdot{\vb*{I}}_k,\,\,B\approx-\frac{1}{2}\sum_k(\vb*{h}_k\cdot{\vb*{I}}_k+2\omega_0 I_k^z).
\end{equation}
Since the nuclear spins are separated here, we can treat each nuclear individually, we denote $A_k=-\frac{1}{2}\vb*{h}_k\cdot{\vb*{I}}_k=\vb*{a}_k\cdot \vb*{\sigma}_k/2$ and $B_k=-\frac{1}{2}(\vb*{h}_k\cdot{\vb*{I}}_k+2\omega_0 I_k^z)=\vb*{b}_k\cdot \vb*{\sigma}_k/2$.

We first consider a single nuclear spin, and omit the subscript $k$ for simplicity. The noise operator in the interaction picture is
\begin{equation}
\begin{aligned}
        A_I(t)&=e^{i\vb*{b}\cdot \vb*{\sigma}t/2}\vb*{a}\cdot \vb*{\sigma}e^{-i\vb*{b}\cdot \vb*{\sigma}t/2}\\
        &=[\vb* a\cos(\omega t)+\hat{\vb*{b}}\times \vb*{a}\sin(\omega t)+\hat{\vb*  b}\hat{\vb*  b}\cdot\vb* a(1-\cos(\omega t))]\cdot\vb*\sigma,
\end{aligned}
\end{equation}
here we let $\vb* b=\omega\hat{\vb*{b}}$ with $\hat{\vb*{b}}$ being a unit vector and $\omega=\frac{1}{2}\sqrt{(h^x)^2+(h^y)^2+(h^z+2\omega_0)^2}$ being the effective Larmor frequency, we use $e^{i\frac{b}{2}(\hat{\vb*  n}\cdot\vb*\sigma)}\vb*{\sigma}e^{-i\frac{b}{2}(\hat{\vb*  n}\cdot\vb*\sigma)}=\vb*\sigma\cos(b)+\hat{\vb*  n}\times \vb*{\sigma}\sin(b)+\hat{\vb*  n}\hat{\vb*  n}\cdot\vb*\sigma(1-\cos(b))$. Then
\begin{equation}
\begin{aligned}
        A_I(t)A_I(0)&=\{[\vb* a\cos(\omega t)+\hat{\vb*{b}}\times \vb*{a}\sin(\omega t)+\hat{\vb*  b}\hat{\vb*  b}\cdot\vb* a(1-\cos(\omega t))]\cdot\vb*\sigma\}(\vb*{a}\cdot \vb*{\sigma})\\
        &=[a^2\cos(\omega t)+(\vb* b\cdot\vb* a)^2(1-\cos (\omega t))]\mathbb{I}+i(...)\cdot \vb*{\sigma}.
\end{aligned}
\end{equation}
where we use $(\vb*{p}\cdot \vb*{\sigma})\cdot(\vb*{q}\cdot \vb*{\sigma})=(\vb*{p}\cdot\vb*{q})\mathbb{I}+i(\vb*{p}\times\vb*{q})\cdot\vb*{\sigma}$ in the above function. Similarly,
\begin{equation}
\begin{aligned}
    A_I(0)A_I(t)&=(\vb*{a}\cdot \vb*{\sigma})\{[\vb* a\cos(\omega t)+\hat{\vb*{b}}\times \vb*{a}\sin(\omega t)+\hat{\vb*  b}\hat{\vb*  b}\cdot\vb* a(1-\cos(\omega t))]\cdot\vb*\sigma\}\\
        &=[a^2\cos(\omega t)+(\vb* b\cdot\vb* a)^2(1-\cos (\omega t))]\mathbb{I}-i(...)\cdot \vb*{\sigma}
\end{aligned}
\end{equation}
We suppose that the bath is in the maximally mixed state, then the second term does not contribute to the correlation function, and the noise correlation function of a single spin becomes
\begin{equation}
\begin{aligned}
    \mcc(t)&=\Tr\left\{\rho[  A_I(t) A_I(0)+ A_I(0) A_I(t)]\right\}/2\\
    &=\Tr{[a^2\cos(\omega t)+(\vb* b\cdot\vb* a)^2(1-\cos (\omega t))]\mathbb{I}}/2\\
    &=|\vb*a_\parallel|^2+|\vb*a_\perp|^2\cos(\omega t)\\
    &=\frac{1}{4}[|\vb*h_\parallel|^2+|\vb*h_\perp|^2\cos(\omega t)],
\end{aligned}
\end{equation}
where $\vb*a_\parallel=\vb*a(\vb*a\cdot\hat{\vb*b})$ and $\vb*a_\perp=\vb*a-\vb*a_\parallel$. 

Then for the whole spin bath, the noise correlation function is
\begin{equation}
\mcc(t)=\frac{1}{4}\sum_k[|\vb*h_{k,\parallel}|^2+|\vb*h_{k,\perp}|^2\cos(\omega_k t)],
\end{equation}
and the noise spectrum becomes
\begin{equation}
    S(\omega)=\frac{\pi}{2}\sum_k[|\vb*h_{k,\parallel}|^2\delta(\omega)+\frac{1}{2}|\vb*h_{k,\perp}|^2(\delta(\omega-\omega_k)+\delta(\omega+\omega_k))],
\end{equation}
which shows peaks at the effective Larmor efficiencies $\omega_k$ with heights proportional to the transverse component of the hyperfine coupling $\vb*h_{k,\perp}$.

\subsection{Noise spectroscopy of nuclear spin clusters}

Here we extend our analysis to a nuclear spin cluster, where the dipolar couplings are no longer neglectable. Then we choose the subspace $\{\ket{m_s=1}_q,\ket{m_s=-1}_q\}$, and the Hamiltonian can be recast to
\begin{equation}
    P_{+1}HP_{+1}+P_{-1}HP_{-1}=\sigma^z_q\otimes \sum_k\vb*{h}_k\cdot{\vb*{I}}_k-\omega_0\sum_k I_k^z+\sum_{j<k}D_{jk}'\big[{\vb*{I}}_j\cdot{\vb*{I}}_k-3({\vb*{I}}_j\cdot\vb*{r}_{jk})({\vb*{I}}_k\cdot\vb*{r}_{jk})/r_{jk}^2\big],
\end{equation}
then
\begin{equation}
    A=\sum_k\vb*{h}_k\cdot{\vb*{I}}_k,\quad B=-\omega_0\sum_k I_k^z+\sum_{j<k}D_{jk}'\big[{\vb*{I}}_j\cdot{\vb*{I}}_k-3({\vb*{I}}_j\cdot\vb*{r}_{jk})({\vb*{I}}_k\cdot\vb*{r}_{jk})/r_{jk}^2\big].
\end{equation}
In a strong magnetic field, the Hamiltonian can be further approximated as
\begin{equation}
    A\approx \sum_k h_k^z I_k^z,\quad B\approx-\omega_0\sum_k I_k^z +\sum_{j<k}D_{jk}[I_j^zI_k^z-(I_j^xI_k^x+I_j^yI_k^y)/2],
\end{equation}
where $D_{jk}=D_{jk}'(1-3\cos^2\theta_{jk})$ with $\cos\theta_{jk}={\vb* r_{jk}\cdot \hat{\vb* z}}/{|\vb* r_{jk}|}$. 

During the free evolution period, the NV electron is projected into the subspace $\ket{m_s=0}_q$, then the Hamiltonian becomes
\begin{equation}
    P_0HP_0=-\omega_0\sum_k I_k^z +\sum_{j<k}D_{jk}[I_j^zI_k^z-(I_j^xI_k^x+I_j^yI_k^y)/2]=B,
\end{equation}
which is the effective evolution Hamiltonian.

\section{Effect of probe-environment interaction during the free evolution}

\begin{figure}
    \centering
    \includegraphics[width=0.5\linewidth]{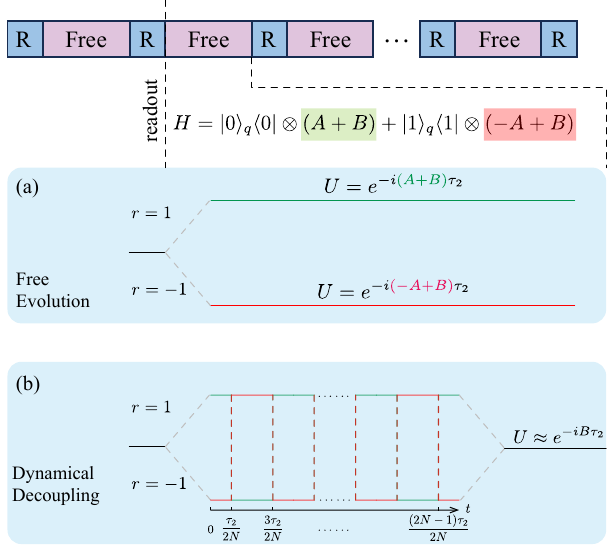}
    \caption{Illustration of probe-bath coupling effects with (a) free evolution and (b) DD. (a) For the free evolution case (no action is performed after readout), the evolution is conditioned on the measurement result, the effective propagator is $U=e^{-i(rA+B)\tau_2}$ with $r=\pm1$. (b) For the DD case, where we consider $N$-pulse CPMG control, the probe is flipped on $t=(2n-1)\tau_2/(2N)$ with $n=1,...,N$. The probe is approximately decoupled with the environment, resulting in the effective propagator $U\approx e^{-iB\tau_2}$.}
    \label{fig:RIMCPMG}
\end{figure}
In this section, we analyze the dephasing effect of probe-environment interaction during the free evolution and introduce a DD-based method to mitigate it.
After each RIM, the readout result of the probe is either $r=1$ or $r=-1$, meanwhile, the probe is projected to $\ket{0}_q$ or $\ket{1}_q$. We can rewrite the Hamiltonian in a probe state dependent form
\begin{equation}
    H=\ket{0}_q\bra{0}\otimes (A+B)+\ket{1}_q\bra{1}\otimes (-A+B).
\end{equation}
 Thus the bath undergoes different evolution trajectories with effective Hamiltonian $H= rA+B$ for $r=\pm 1$ [Fig.~\ref{fig:RIMCPMG}\blue{(a)}]. Since that $[A,B]\neq 0$, the effective propagator $U_r=e^{-i (rA+B)\tau_2}$ does not commutate with $U_B'$, which can cause additional dephasing \cite{Jiang2008}.

Here we introduce the dynamical decoupling (DD) method into the free evolution period to (approximately) eliminate the dephasing in this period. We consider the $N-$pulse Carr-Purcell-Meiboom-Gill (CPMG) control, in which N $\pi$-pulses are applied on the time $t=(2n-1)\tau_2/(2N)$ to flip the probe ($N$ is even here). The the propagator becomes
\begin{equation}
    U_r=(e^{-i(rA+B)\tau_2/2N}e^{-i(-rA+B)\tau_2/N}e^{-i(rA+B)\tau_2/2N})^{N/2}.
\end{equation}
Up to the first order Magnus expansion, we have \cite{Yang2011}
\begin{equation}
    U_{+1}\approx U_{-1} \approx e^{-iB\tau_2}.
\end{equation}

\begin{figure}
    \centering
    \includegraphics[width=1\linewidth]{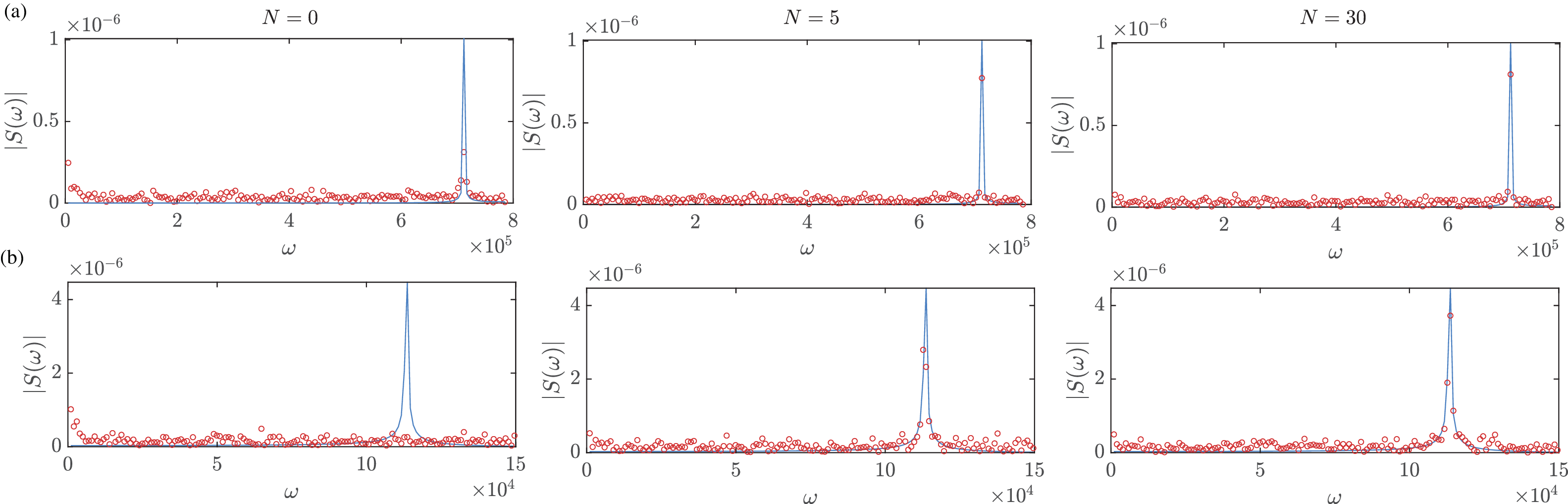}
    \caption{Numerical simulation of dephasing induced by probe–bath coupling and the effect of DD control. The blue solid lines represent the exact noise spectra (multiplied by a small global factor $4\tau_1^2$ ) while the red circles represent the direct noise spectroscopy results obtained from Monte Carlo simulations of repetitive RIMs (with $4\times 10^6$ samples). The peak is suppressed due to the additional dephasing without DD ($N=0$). As the number of DD pulses increases ($N=5$ and $N=30$), the probe–bath coupling is averaged out and the peak is recovered. The parameters are $h=18.2$ kHz, $B=10^{-2}$ T, $\tau_1=1\,\,\mu$s, (a) $\tau_2=3\,\,\mu$s, (a) $\tau_2=20\,\,\mu$s. }
    \label{fig:DDsim}
\end{figure}

\begin{figure}
    \centering
    \includegraphics[width=0.8\linewidth]{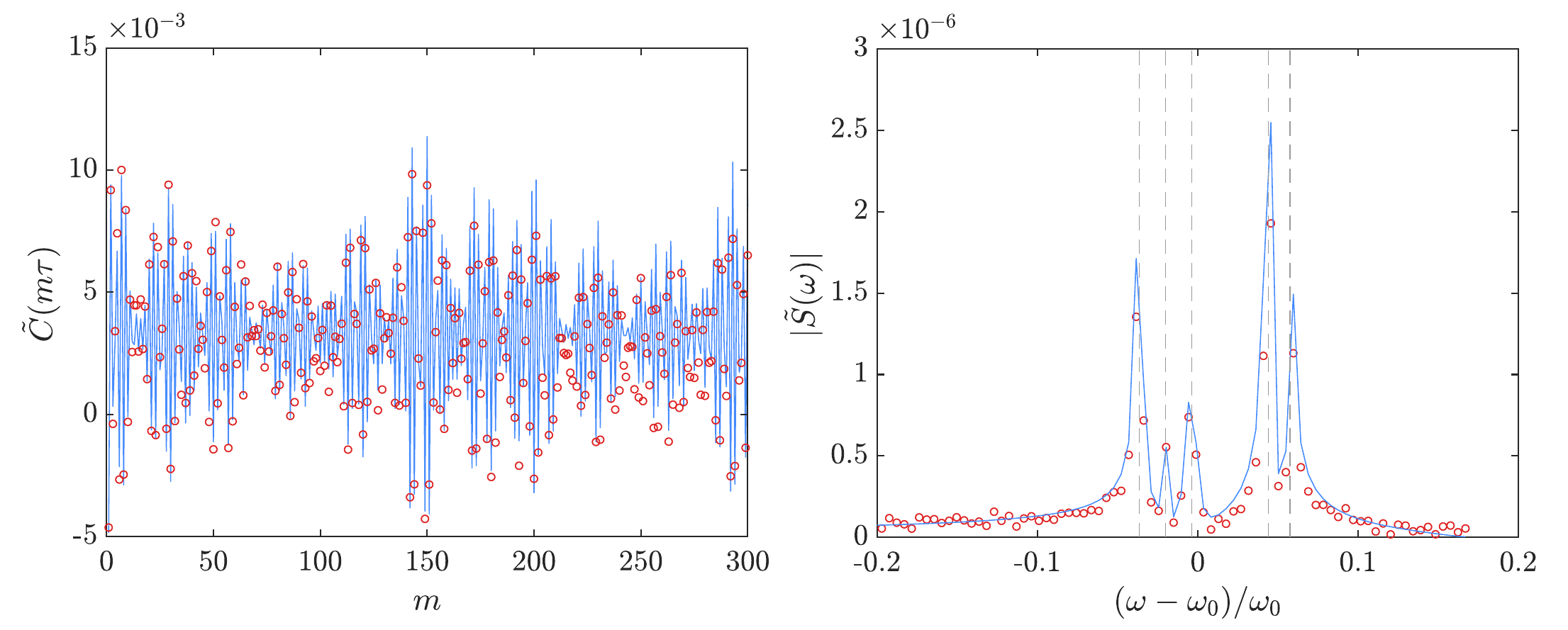}  
    \caption{Direct qubit noise spectroscopy for a spin bath under a DD control. The blue solid lines represent the exact noise spectra (multiplied by a small global factor $4\tau_1^2$ ) while the red circles represent the direct noise spectroscopy results obtained from Monte Carlo simulations of repetitive RIMs (with $4\times 10^6$ samples). The dashed lines in the spectrum indicates the effective Larmor frequencies. The parameters for the spin bath contains five nuclear spins in the maximally mixed state with $\tau_1=1\,\,\mu$s, $\tau_2=3\,\,\mu$s, $B=100$ G, and $\{|\vb*h_k|\}_{k=1}^5=\{0.105,0.113,0.103,0.107,0.112\}$ MHz.}
    \label{fig:fivespinsim}
\end{figure}

We show the effect of dephasing and DD in a single-qubit target system [Eq.~\eqref{Eq:Ham0-1} with only one nuclear spin] in Fig.~\ref{fig:DDsim}. We show the case for short evolution time [Fig.~\ref{fig:DDsim}\blue{(a)}] and long evolution time [Fig.~\ref{fig:DDsim}\blue{(b)}]. We can see that the peak shows a broadening without the DD control ($N=0$), and is recovered with DD pulses applied.

Then we can perform direct qubit noise spectroscopy in an environment that cannot be decoupled with the probe, such as Eq.~\eqref{Eq:Ham0-1}, with the help of DD. Here we show a simulation for a dilute spin bath containing five $^{13}$C nuclear spins, which are weakly coupled to each other ($D_{jk} \sim 10\,\text{Hz} \ll h_k \sim 100\,\,\rm{kHz}$). As shown in Fig.~\ref{fig:fivespinsim}, the noise spectroscopy results well reproduce the exact noise spectrum, revealing five peaks centered at the effective Larmor frequencies.

\section{Performance analysis and comparison}
In this section, we consider the performance of our method in detail. First, we analyze the detectable spectral bandwidth based on the sampling theorem. Then we calculate the sample complexity for our method using Hoeffding's inequality. Besides, we show that the noisy environment with Lindblad dissipation can induce additional damping on the correlation function, resulting in the peak broadening of noise spectrum. We also show the resource-efficiency of our method compared to the correlation spectroscopy with different measurement strength. 
\subsection{Detectable spectral bandwidth}
\textbf{Theorem} 1 (Nyquist–Shannon sampling theorem) If a function $x(t)$ contains no frequencies higher than 
$\Omega$, then it is completely determined by its sampled values 
$x(m\tau)$ with $m=1,2,...,N$ satisfying $\tau\leq\frac{\pi}{\Omega}$. The frequency resolution is $\Delta\omega=\pi/(N\tau)$.

Then, for a certain sampling duration $\tau$, the frequency bandwidth we can detect becomes 
\begin{equation}\label{omegaregion}
    \omega\in[0,\pi/\tau].
\end{equation} Since the frequencies $\omega_{ij}=b_i-b_j\leq 2||B||$, then the limitation for the total evolution time is $\tau\leq\pi/(2||B||)$, with no requirement on the relative magnitude of $||A||$ and $||B||$. In addition, according to the expansion in  Eq.~\eqref{U0U1}, $\tau_1||B||$ cannot be too large so that higher-order terms will be induced although we do not need $\tau_1||B||\ll 1$.

\begin{figure}[htbp]
    \centering
    \includegraphics[width=\linewidth]{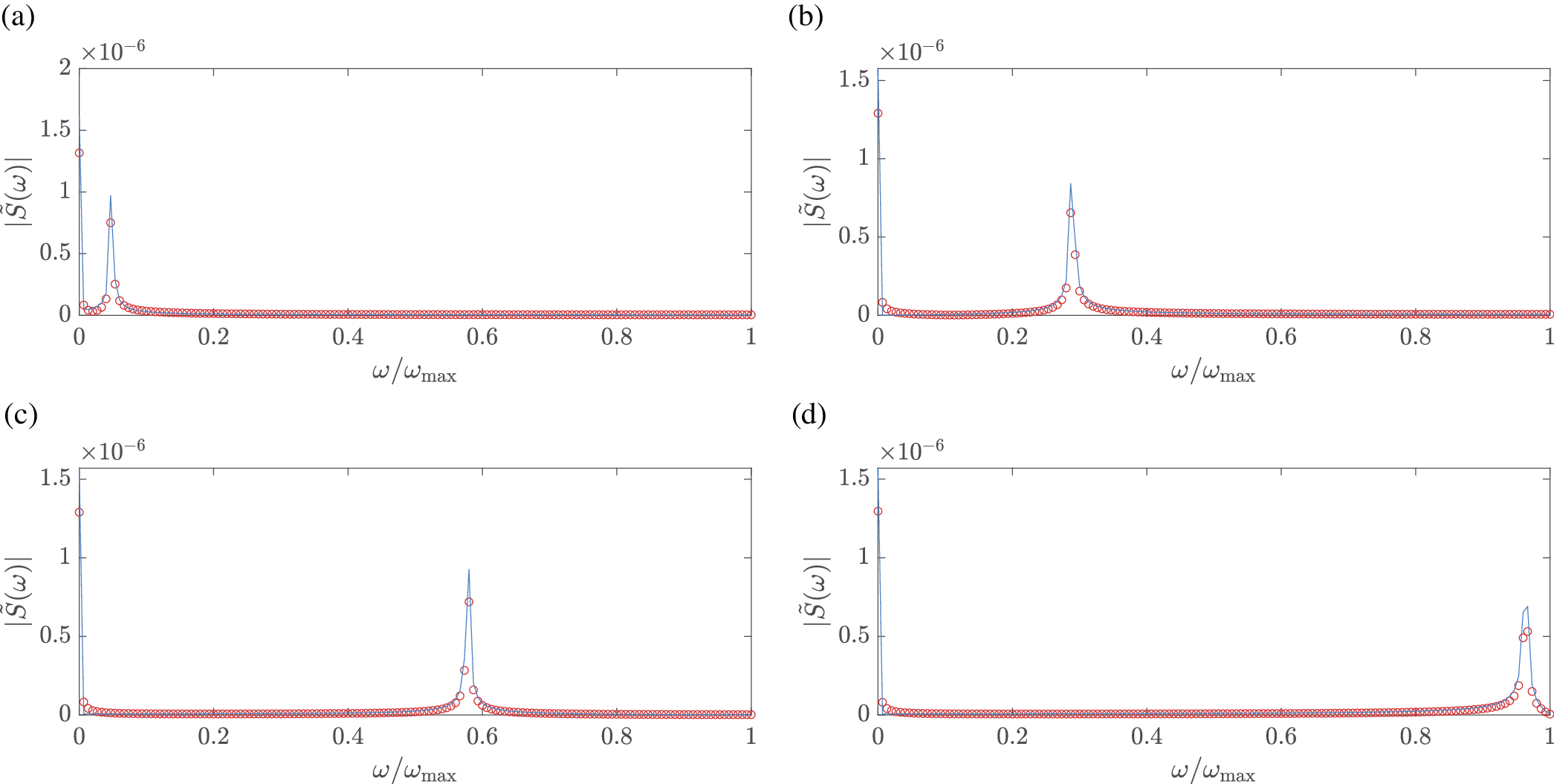}
    \caption{Numerical simulation of direct noise spectroscopy via sequential weak measurements with different signal frequencies. The parameters are $\tau_1=0.6\,\,\mu$s, $\tau=2.6\,\,\mu$s (i.e., $\omega_{\max}= 1.93\times 10^5$ Hz) with (a) $||B||/||A||=0.508$, (b) $||B||/||A||=3.05$, (c) $||B||/||A||=6.09$ and (d)$||B||/||A||=10.2$.}
    \label{fig:signalfreq}
\end{figure}

Below we numerically demonstrate that the noise spectrum can be accurately reconstructed within the frequency window defined in Eq.~\eqref{omegaregion} for a single-qubit environment [Fig.~\ref{fig:signalfreq}]. 

\subsection{Sample complexity}
Here we use the Hoeffding's inequality to derive the sample requirement of reconstructing the signal.

\textbf{Theorem} 2 (Hoeffding's inequality) Let $Z_1,...,Z_{N_s}$ be independent bounded random variables with $Z_i\in [a, b]$ for all $i$, where $-\infty<a<b<\infty$, then
\begin{equation}
    \mathrm{Pr}\left(\frac{1}{N_s}\abs{\sum_{i=1}^{N_s}(Z_i-\mathbb{E}[Z_i])}\geq \delta\right)\leq2\exp(-\frac{2N_s\delta^2}{(b-a)^2}).
\end{equation}
For each sample, the obtained two-point correlation is $r_1 r_m\in\{-1,1\}$, and the total correlation function is estimated as $\hat C^{\rm weak}(m\tau)=\frac{1}{N_s}\sum_{i=1}^{N_s} r_1 r_m$. Then we have
\begin{equation}
     \mathrm{Pr}\left(\abs{\hat C^{\rm weak}(m\tau)- \tilde C^{\rm weak}(m\tau)}\geq \delta\right)\leq2\exp(-\frac{N_s\delta^2}{2}).
\end{equation}
Therefore, to guarantee
\begin{equation}
    \Pr\left(\abs{\hat C^{\rm weak}(m\tau)- \tilde C^{\rm weak}(m\tau)}\leq \delta\right)\geq 1-\varepsilon,
\end{equation}
the sample number should suffices
\begin{equation}
    N_s \geq \frac{2}{\delta^2}\ln(\frac{2}{\varepsilon}).
\end{equation}
For the weak-measurement based methods (including the correlation spectroscopy and our method), the signal amplitude of the correlation function is about $10^{-3}\sim 10^{-2}$, then about $10^6$ samples are needed to reconstruct the correlation function.

\subsection{Spectroscopy for noisy environments}
\begin{figure}[htbp]
    \centering
    \includegraphics[width=0.8\linewidth]{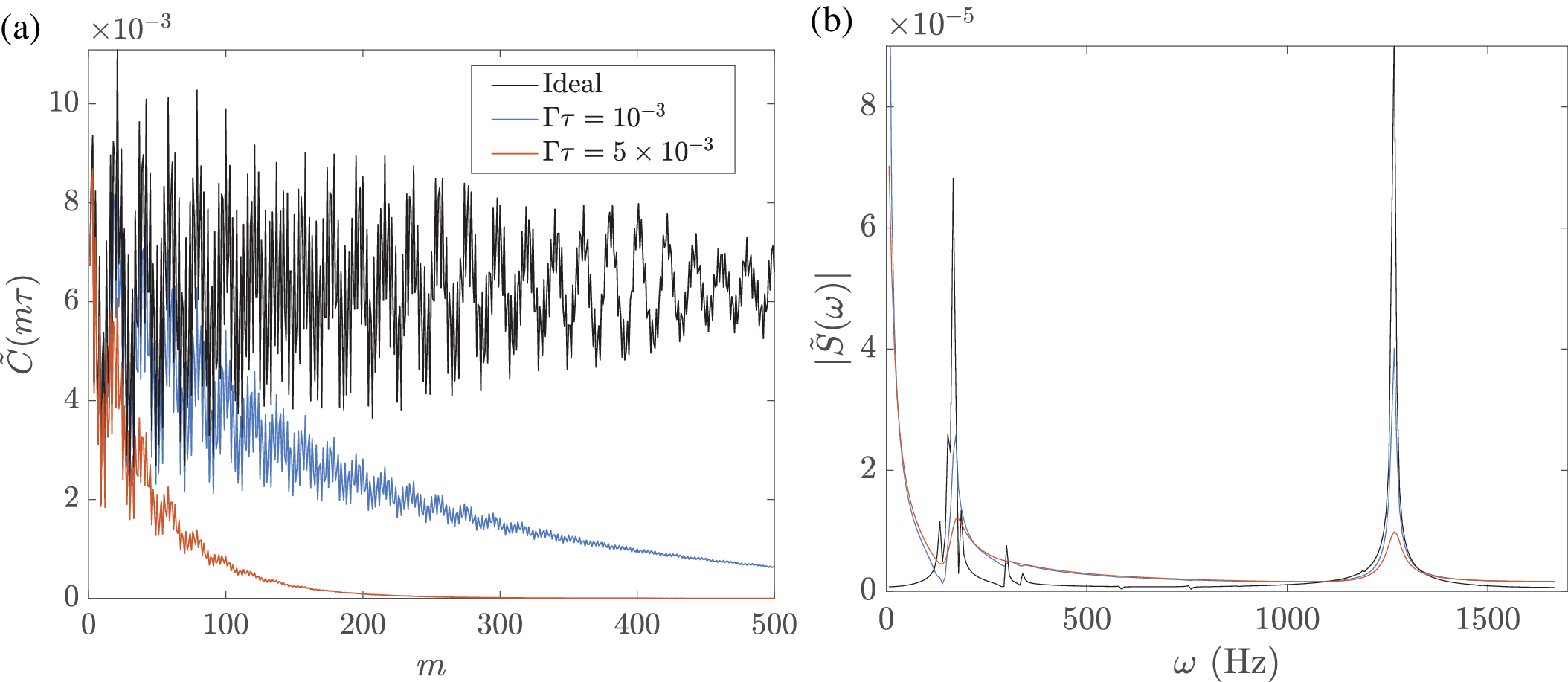}
    \caption{Direct qubit noise spectroscopy in a dissipative environment.  
    (a) Noise correlation function for three dissipation strengths: the ideal correlation function (black line), correlation function measured in our method with $\Gamma\tau = 10^{-3} $ (blue line), and $\Gamma\tau = 5\times 10^{-3}$ (red line).(b) Corresponding noise spectra $|S(\omega)|$ reconstructed from the data in (a).
    }
    \label{fig:noise}
\end{figure}

Here we analyze the performance of noise spectrum extraction for a noisy environment. We use the central spin model in Sec.~\ref{Sec:centralSpin} for illustration, in which we consider the nuclear spin relaxation and dephasing. The evolution of the system can be described by the Lindblad master equation
\begin{equation}
    \dv{\rho_{\rm tot}}{t}=-i[H,\rho_{\rm tot}]+\sum_{k}\mathcal{D}[\sqrt{\Gamma_1}\sigma_k^-](\rho_{\rm tot})+\sum_{k}\mathcal{D}[\sqrt{\Gamma_\phi}\sigma_k^z](\rho_{\rm tot})
\end{equation}
where $\rho_{\rm tot}$ is the state of the composite system, $\Gamma_\phi$ is the dephasing rate, $\Gamma_1$ is the relaxation rate, and $\mathcal{D}[X](\cdot)=X(\cdot) X^\dagger-\frac{1}{2}\{X^\dagger X,(\cdot)\}$ is the Lindblad dissipator with $X$ being the jump operator and $\{X,Y\}=XY+YX$ denoting the anti-commutator. We perform numerical simulation to show the influence in Fig. \ref{fig:noise}. We can see that the dissipation can induce additional damping onto the correlation function [Fig. \ref{fig:noise} \blue{(a)}], which broadens the peaks in the noise spectrum [Fig. \ref{fig:noise} \blue{(b)}].

\subsection{Comparison of performance with different measurement strength}

\begin{figure}[htbp]
    \centering
    \includegraphics[width=\linewidth]{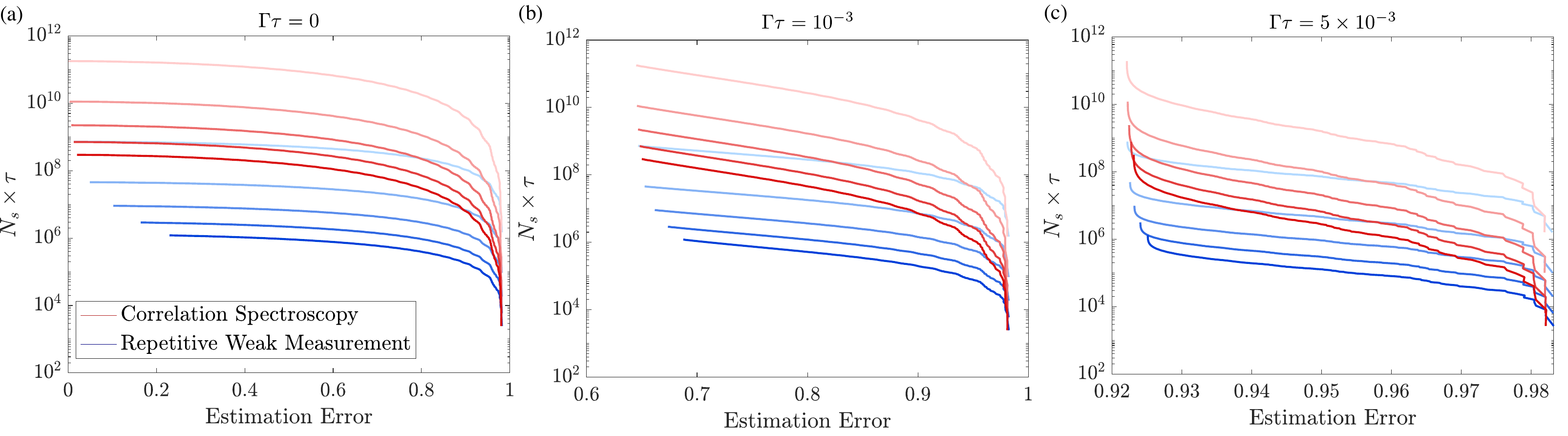}
    \caption{Comparison on resource complexity of correlation spectroscopy and our method with (a) $\Gamma\tau=0$, $\Gamma\tau=10^{-3}$, and $\Gamma\tau=5\times10^{-3}$. The color shading represents different measurement strengths, with darker colors corresponding to stronger measurements (realized by longer $\tau_1$). }
    \label{fig:accuracydiss}
\end{figure}

 The accuracy of correlation spectroscopy is not limited by the broadening of peaks caused by the damping factor related to the measurement strength. Then stronger measurement strength (longer $\tau_1$) can be taken in that method to obtain a higher signal amplitude, and thus the sample complexity can be lower. Here we numerically compare the resource complexity (defined as $N_s\times t_{\rm tot}$) of correlation spectroscopy and our method in different measurement and noise strengths (Fig.~\ref{fig:accuracydiss}). We can see that our approach demonstrates advantages within a certain range of estimation accuracy, and these advantages are further enhanced in noisy practical systems.

\bibliography{ref}